\documentclass[twocolumn,prl,superscriptaddress]{revtex4}

\usepackage{graphicx}
\usepackage{dcolumn}
\usepackage{bm}
\usepackage{bbm}

\begin{document}

\title{Universal Long-time Behavior of Nuclear Spin Decays in a Solid}

\author{S.\ W.\ Morgan}
\altaffiliation[Present address: ]{Department of Physics, Princeton
University, Princeton, NJ 08544, USA} \affiliation{Department of
Physics, University of Utah, 115 South 1400 East, Salt Lake City,
Utah 84112-0830, USA}
\author{B.\ V.\ Fine}
\affiliation{Department of Physics and Astronomy, University of
Tennessee, Knoxville, TN 37996-1200, USA} \affiliation {Institut
f\"{u}r Theoretische Physik, Universit\"{a}t Heidelberg,
Philosophenweg 19, 69120 Heidelberg, Germany}
\author{B.\ Saam}
\email{saam@physics.utah.edu} \affiliation{Department of Physics,
University of Utah, 115 South 1400 East, Salt Lake City, Utah
84112-0830, USA}

\date{April 30, 2008}

\begin{abstract}
Magnetic resonance studies of nuclear spins in solids are
exceptionally well suited to probe the limits of statistical
physics. We report experimental results indicating that isolated
macroscopic systems of interacting nuclear spins possess the
following fundamental property: spin decays that start from
different initial configurations quickly evolve towards the same
long-time behavior. This long-time behavior is characterized by the
shortest ballistic microscopic timescale of the system and therefore
falls outside of the validity range for conventional approximations
of statistical physics. We find that the nuclear free induction
decay and different solid echoes in hyperpolarized solid xenon all
exhibit sinusoidally modulated exponential long-time behavior
characterized by identical time constants. This universality was
previously predicted on the basis of analogy with resonances in
classical chaotic systems.
\end{abstract}

\maketitle

The relationship between statistical physics and chaos is one of the
most important and controversial problems in theoretical physics.
Statistical physics is based on the assumption of some kind of
randomness on the microscopic scale, yet the question of whether
this randomness is at all related to the mathematical concept of
chaos (well-established for few-body classical systems) is not well
understood \cite{gaspardbook,haakebook}. In many-body systems, it is
extremely difficult to separate the effects of randomness associated
with true chaos from those associated with averaging over the
macroscopic number of degrees of freedom
\cite{gaspard98,dettmann99,grassberger99,gaspard99}. The situation
is further complicated by the lack of consensus on the universal
definition of chaos in quantum systems. In view of these
complications, one approach is to proceed on the basis of
conjectured parallels between the properties of mathematical chaotic
systems and real many-body systems. The predicted consequences of
these conjectures can then be tested numerically or experimentally.
One such prediction about the universal long-time behavior of
transient nuclear spin decays in solids has been made recently in
Ref.~\cite{fine05}. The work presented here tested that prediction
by measuring the transverse relaxation of $^{129}$Xe nuclei (spin =
1/2) in solid xenon over four orders of magnitude using nuclear
magnetic resonance (NMR). Such experiments are prohibitively
challenging for conventional NMR due to the weak thermal
magnetization achievable in even the strongest magnets. We have
employed the technique of spin-exchange optical pumping
\cite{walker97} in order to achieve enhanced (hyperpolarized)
magnetization required for this experiment.

In nearly perfect agreement with the prediction of
Ref.~\cite{fine05}, our experiments indicate that the long-time
behavior of transverse nuclear spin decays in solids has the
universal functional form
\begin{equation}
\label{eq:lt} F(t) = A{\textrm{e}}^{-\gamma t}\cos(\omega t + \phi),
\end{equation}
where the decay coefficient $\gamma$ and the beat
frequency $\omega$ are {\it independent of the initially generated
transverse spin configuration}. This long-time behavior sets in
after only a few times $T_2$, where $T_2$ is the characteristic
timescale for transverse decay determined by the interaction between
nuclear spins (see Eq.~(\ref{eq:dipdip}) below) and represents the
shortest ballistic timescale in the system. The values of $1/\gamma$
and $1/\omega$ are also on the order of $T_2$. Hence, it cannot be
that the spins are interacting with a fast-equilibrating heat bath,
which would justify the exponential character of the decay, as for a
common damped harmonic oscillator. Indeed, at 77~K in an applied
magnetic field $\geq 1$~T, the $^{129}$Xe spins are well isolated
from their environment. The longitudinal relaxation time $T_1
\approx 2.3$~h \cite{gatzke93} while $T_2 \approx 1$~ms
\cite{yen63}; therefore, the decay cannot be attributed to
spin-lattice relaxation. The oscillations in this decay, sometimes
referred to as Lowe beats \cite{lowe57}, constitute a correlation
effect \cite{fine04,fine97} induced by the spin-spin interaction and
have nothing to do with the Larmor frequency. We verified that the
effects of radiation damping and inhomogeneities in the external
field are also negligible on the timescale of $T_2$ \cite{epaps}.
The observed decay thus represents the approach of a closed quantum
system to equilibrium.

\begin{figure}
\begin{center}
\includegraphics[width=3in]{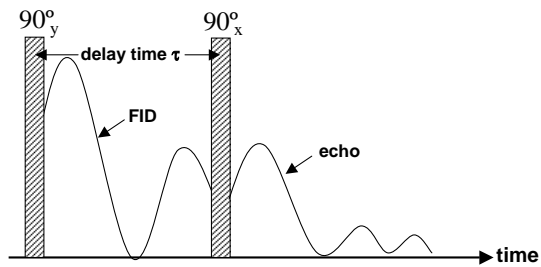}
\caption{\label{fig:seq} The pulse sequence used to generate a solid
echo. The magnitude of the free-induction decay (FID) including Lowe
beats is shown schematically after the first pulse with the
solid-echo response shown after the second pulse. The pulses are
separated in phase by 90$^{\circ}$: the first is along the
rotating-frame y-axis, the second is along the rotating-frame
x-axis. Unlike conventional Hahn echoes, the solid echo does not
generally peak at time $2\tau$\cite{epaps}.}
\end{center}
\end{figure}

We used the particular pulse sequence (see Fig.~\ref{fig:seq}),
known as a solid echo \cite{powles62,epaps}, which consists of two
$90^{\circ}$ pulses (the first along the y-axis and the second along
the x-axis in the rotating frame) separated by a delay time $\tau$.
In contrast with the conventional Hahn spin echo \cite{hahn50} or
the magic echo \cite{rhim71}, the solid echo is not an
amplitude-attenuated reproduction of the free induction decay (FID)
that peaks at time 2$\tau$. Complete refocusing by solid echoes
occurs only for isolated pairs of spins \cite{powles62}. A deviation
from complete refocusing is caused by higher-order correlations
involving more than two spins. The solid echo response depends on
the spin configuration just after the second pulse, whereby
different values of the delay time $\tau$ imply fundamentally
different ``after-pulse" configurations \cite{epaps} that evolve
from the uniformly polarized uncorrelated spin state at the
beginning of the FID to highly correlated states induced by
spin-spin interactions during the delay time \cite{cho05}.
Experimentally, these distinct after-pulse configurations are
exactly what is required in order to clearly demonstrate the
evolution to a universal long-time behavior.

On the timescale of our experiments, the system of interacting
$^{129}$Xe nuclei can be accurately described as isolated and
governed by the Hamiltonian of the truncated magnetic dipolar
interaction \cite{slichterbook,abragam61}, which in the Larmor
rotating reference frame has the form
\begin{equation}
\label{eq:dipdip} {\mathcal{H}} =\sum_{k<n}
[B_{kn}(I_{k}^{x} I_{n}^{x} +
I_{k}^{y} I_{n}^{y})+A_{kn} I_{k}^{z} I_{n}^{z}],
\end{equation}
where $A_{kn}$ and $B_{kn}$ are coupling constants and $I_{n}^{i}$
are the spin operators representing the $i$th projection of the
$n$th spin. The Hamiltonian in Eq.~(\ref{eq:dipdip}) is appropriate
in the high-field limit where the Zeeman energy dominates dipolar
couplings; hence, the shape and duration of the FID are independent
of the applied field. The characteristic decay timescale $T_2$ is on
the order of a few inverse nearest-neighbor coupling constants. Although
the coupling constants in the Hamiltonian Eq.~(\ref{eq:dipdip}) can
be very accurately determined from first principles, efforts over
several decades
\cite{lowe57,slichterbook,abragam61,vvleck48,klauder62,powles63,borckmans68,cowanbook}
to predict the entire behavior of FIDs and spin echoes
quantitatively have met only with limited success; direct
calculations tend to lose predictive power in the long-time tail of
the decay, where increasingly higher-order spin correlations become
important \cite{cho05}. Although methods have not yet been developed
for the controllable calculation of $\omega$ and $\gamma$ in
Eq.~(\ref{eq:lt}), Ref.~\cite{fine05} predicted the quick onset
\cite{fine04} of the long-time behavior of Eq.~(\ref{eq:lt}) with
the same values of $\omega$ and $\gamma$ for all kinds of transverse
decays in the same system. This prediction was based not on a
conventional statistical theory but on a conjecture \cite{fine04}
that quantum spin dynamics generates extreme randomness analogous to
classical chaos.

The long-time behavior of Eq.~(\ref{eq:lt}) for the FID alone
has been previously observed in NMR experiments on $^{19}$F in
CaF$_{\textrm{2}}$ \cite{enge74}. Analogous observations have also
been made in numerical simulations of both classical \cite{fine03}
and quantum \cite{fabricius97} spin lattices. Our experiment adds a
new insight into this universality by demonstrating that the above
long-time behavior is common to both FIDs and solid echoes in the
same spin system. Since solid echoes initiated at different delay
times $\tau$ start from distinct initial spin configurations, our
findings suggest that the tails of transient nuclear spin signals
are independent of initial conditions (apart from the oscillation
phase and the overall amplitude).

For our experiments, both isotopically natural ($26.4\%$ $^{129}$Xe,
$21.29\%$ $^{131}$Xe) and enriched ($86\%$ $^{129}$Xe, $ 0.13\%$
$^{131}$Xe) samples of solid polycrystalline xenon containing
$^{129}$Xe polarized to 5-10\% were prepared using spin-exchange
convection cells \cite{su04,epaps}. FIDs and solid echoes were
acquired at 77~K in an applied field of 1.5~T ($^{129}$Xe Larmor
frequency of 17.6~MHz), well into the high-field limit of
Eq.~(\ref{eq:dipdip}). The enormous dynamic range of these signals
required a separate acquisition of the initial and long-time decays
for each FID and echo using different gain settings for the NMR
receiver \cite{epaps}.

\begin{figure}
\begin{center}
\includegraphics[width=3.0in]{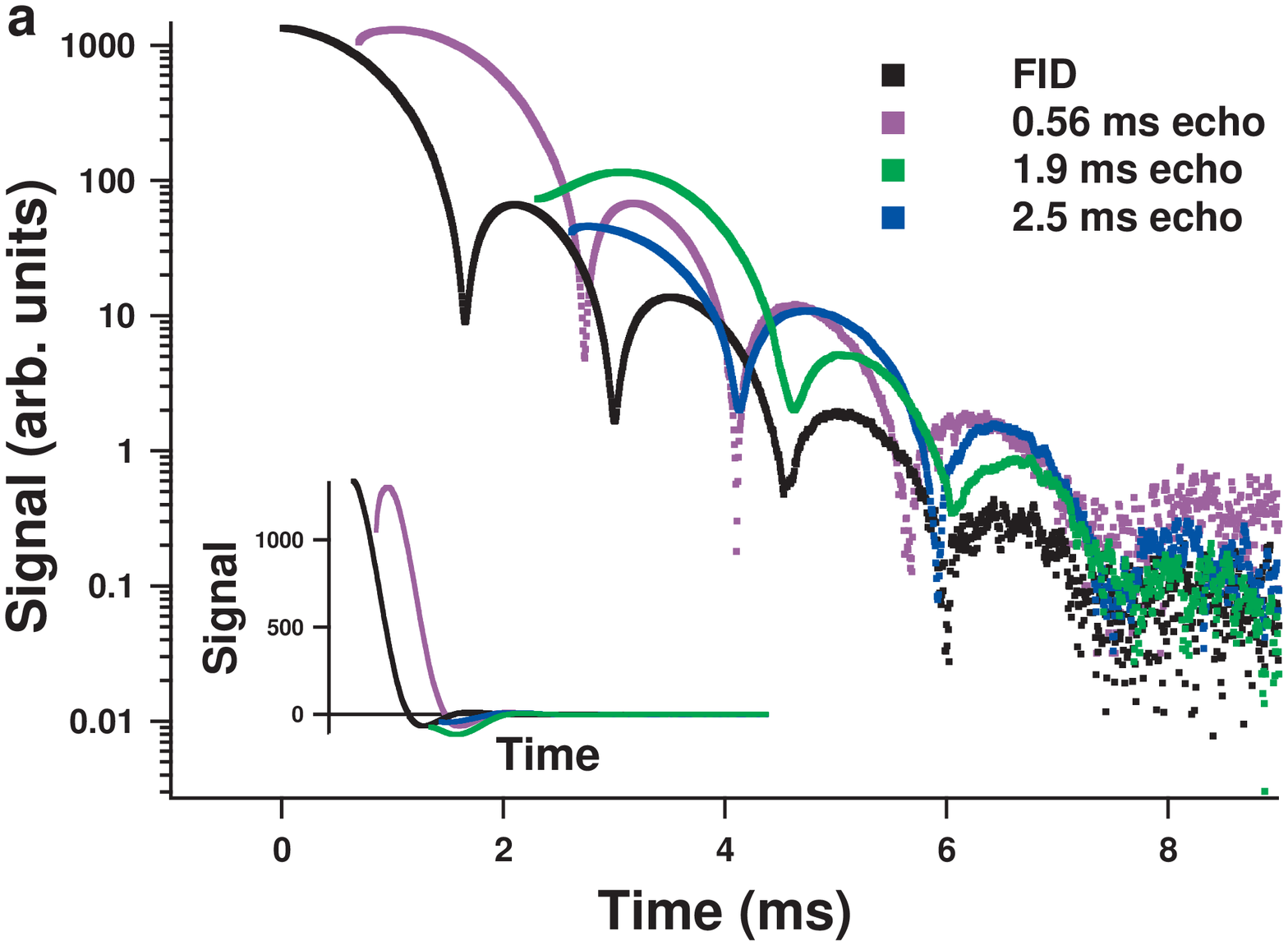}
\includegraphics[width=3.0in]{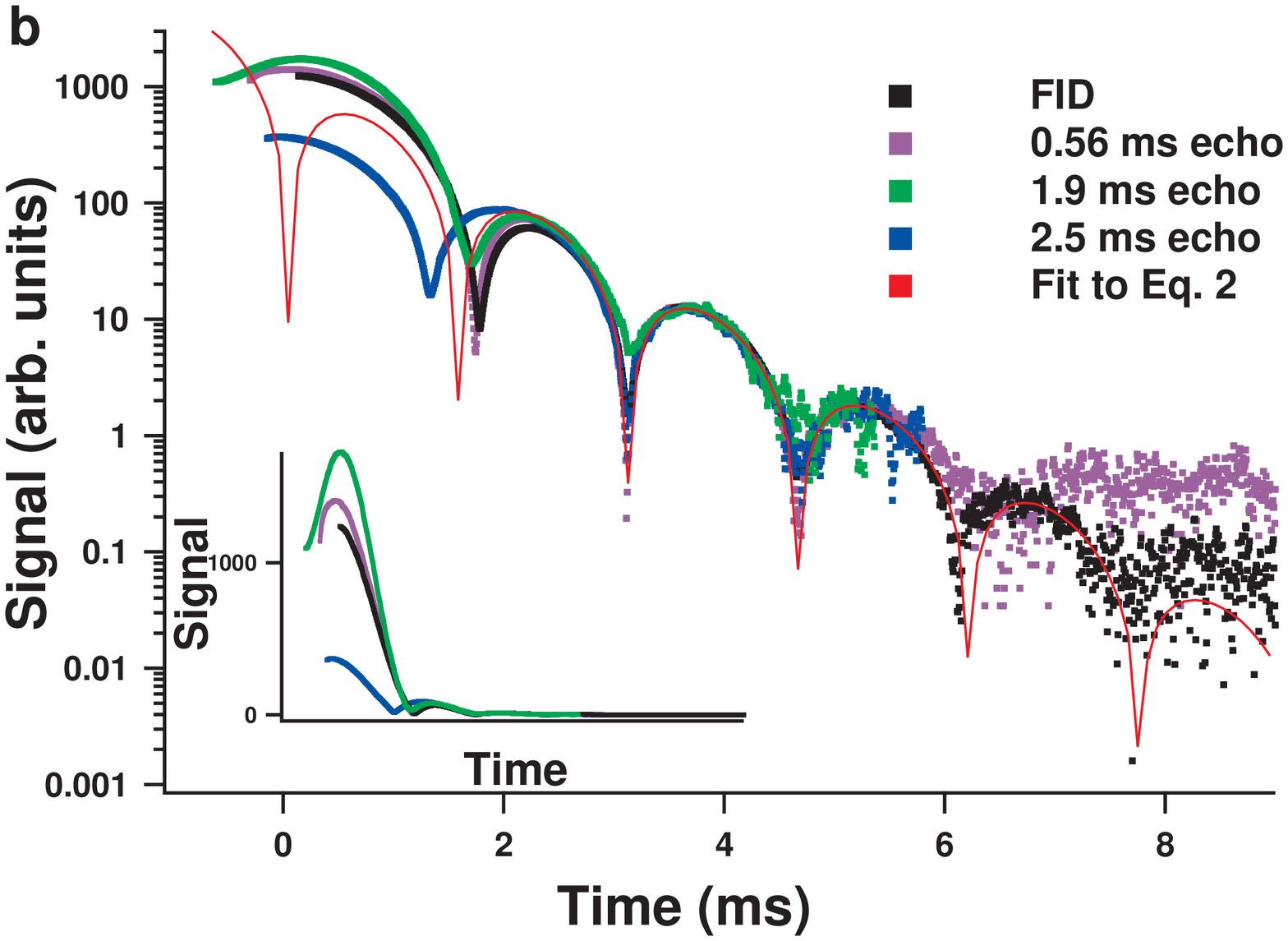}
\caption{\label{fig:decays} Representative acquisitions of FID and
solid echoes (with three different delay times) recorded for
$^{129}$Xe in enriched polycrystalline xenon at 77 K. {\bf a.}\ The
signal magnitudes are shown on a semilog plot (main) and exhibit
characteristic beats. The actual signals plotted on a linear scale
(inset) change sign, whereas the main plot has cusps at the
zero-crossing points. The FID starts at $t = 0$~ms, and each echo
starts at its respective delay time $\tau$ with its initial value
normalized to the value of the FID at time $\tau$. {\bf b.}\ The
same data are shown again on a semilog plot (main) with the echoes
both time- and amplitude-shifted to illustrate the nearly perfect
overlap of the long-time decays. For the 1.9~ms and 2.5~ms echoes,
the noisiest data farthest out in time have been removed for
clarity. The red line is a representative long-time fit to the
absolute value of Eq.~(\ref{eq:lt}) with the decay coefficient
$\gamma = 2.04$~ms$^{-1}$ and the beat frequency $\omega =
1.25$~rad/ms. The distinct differences among the initial portions of
the decays can be better appreciated in the linear absolute-value
plot (inset).}
\end{center}
\end{figure}

In Fig.~\ref{fig:decays}a, representative decays of the signal
magnitude for the FID and solid echoes with three different delay
times $\tau$ are shown for enriched xenon on a semilog plot, with
the time axis referenced to $\approx 100$~$\mu$s (instrumental dead
time) after the end of the first 90$^{\circ}$ pulse, i.e., at the
start of the FID. The FID and each echo are acquired separately with
the sample newly polarized, whereby the run-to-run variation in
polarization prohibits a direct measurement of their relative
amplitudes. Hence, the data for each echo are shown at the proper
temporal location, starting $\approx 100$~$\mu$s after the
corresponding value of $\tau$, and each echo is normalized to match
the FID amplitude at $t = \tau$. Fig.~\ref{fig:decays}b shows the
same four acquisitions time-shifted and amplitude-normalized
relative to the FID to yield the best overlap at long times
($\approx 2.7$~ms and later after the start of the FID or echo). The
decay coefficient $\gamma$ and beat frequency $\omega$ were obtained
for each decay from a fit of the long-time signal magnitude to the
absolute value of Eq.~(\ref{eq:lt}); a representative fit is shown
in red. The results are summarized in Table~\ref{tb:ltfits}, where
each entry represents the average of fits for six separate
acquisitions of the FID or solid echo. For a given isotopic
concentration of $^{129}$Xe, the parameters $\gamma$ and $\omega$
are the same for both the FID and all solid echoes independent of
the delay time $\tau$.

\begin{table}
\begin{center}
\begin{tabular}{|l|l|l|}
\hline
 & $\gamma$ (ms$^{-1}$) & $\omega$ (rad/ms)\\\hline
Enriched FIDs & $1.25 \pm 0.05$&$2.03 \pm 0.04$\\\hline Enriched
echoes, $\tau = 0.56$~ms&$1.25 \pm 0.05$ & $2.00 \pm 0.03$\\\hline
Enriched echoes, $\tau = 1.9$~ms& $1.22 \pm 0.04$ & $2.06 \pm
0.03$\\\hline Enriched echoes, $\tau = 2.5$~ms& $1.25 \pm 0.04$ &
$2.05 \pm 0.04$\\\hline Natural FIDs & $1.04 \pm 0.08$ &$1.53 \pm
0.08$ \\\hline Natural echoes, $\tau = 0.56$~ms& $1.04 \pm 0.12$ &
$1.52 \pm 0.04$
\\\hline
\end{tabular}
\end{center}
\caption[Results of six separate FID experiments and six separate
echo experiments.] {\label{tb:ltfits} The decay coefficient $\gamma$
and beat frequency $\omega$ extracted from the fit of long-time data
by Eq.~(\ref{eq:lt}) for FID and solid echo experiments in both
natural and $^{129}$Xe-enriched solid xenon. Each entry represents
an average of six separate experiments with the errors determined
from the spread in the fit results. The delay time $\tau$ is the
time between the 90$^{\circ}$ pulses in the solid echo pulse
sequence.}
\end{table}

In contrast, there is no universal behavior in the initial portion
of the transverse decays. This is connected to the theoretical
expectation (discussed above) that the longer values of the delay
time $\tau$ allow higher-order spin correlations that are not
refocused by the solid echo to become stronger. As a result, the
initial spin configurations for the FID and various solid echoes are
different and not trivially related to one another \cite{epaps}.

In natural xenon, the long-time tails of the FID and the solid echo
acquired with delay time $\tau = 0.56$~ms are also nearly identical
\cite{epaps} and can be fit by Eq.~(\ref{eq:lt}) with parameters
$\gamma$ and $\omega$ given in Table~\ref{tb:ltfits}. The values of
both of these parameters are smaller than in the enriched sample
because in the natural sample the $^{129}$Xe spins are more dilute,
having been replaced with zero-spin species or with $^{131}$Xe, for
which the dipolar interaction has different coupling constants. The
intrinsically weaker signal meant that the long-time tails of echoes
with delay times $\tau \gtrsim 0.56$~ms could not be accurately
measured.

The common quantitative long-time character of FIDs and spin echoes
provides experimental support for the notion of eigenmodes of
the time evolution operator $\hat{T}(t)$ in isolated many-body
quantum systems. This operator is defined by the equation
$\varrho(t, \mathbf{x}) = \hat{T}(t) \varrho(0, \mathbf{x}) $,
where $\varrho(0, \mathbf{x})$ is the many-body density
matrix at some initial time $t=0$, and $\mathbf{x}$ is the set of
variables that describe the density matrix.  It was conjectured
\cite{fine05,fine04} that in the
observable long-time range, the non-equilibrium behavior of the
density matrix for any small but macroscopic subsystem of the closed
system is controlled by a complex-valued eigenmode having
the form:
\begin{equation}
\varrho_0(\mathbf{x}) {\textrm{e}}^{(- \gamma + i \omega) t} +
\varrho_0^{*}(\mathbf{x}) {\textrm{e}}^{(- \gamma - i \omega) t}.
\label{eq:prsolution}
\end{equation}
If this conjecture is valid, then the long-time decay of
Eq.~(\ref{eq:lt}) represents not just the property of one relaxation
process, such as the FID, but rather an intrinsic property of the
many-body dynamics of the system, and should manifest itself in
numerous other relaxation processes, such as solid echoes with
different delay times $\tau$.

The eigenmodes of the time evolution operator as defined by
Eq.~(\ref{eq:prsolution}) have no direct relation to the eigenvalues
of the Hamiltonian of the many-body system, but rather they are
expected to be counterparts of the Pollicott-Ruelle resonances
\cite{gaspardbook,ruelle86} in classical hyperbolic chaotic systems.
These resonances depend on the rate of probability loss from coarser
to finer partitions of phase space \cite{gaspardbook,haakebook}. In
many-body quantum systems, there should exist an analogous transfer
of spectral weight from lower to higher order quantum correlations
\cite{cho05}.

Quantum analogs of Pollicott-Ruelle resonances have been observed
numerically in kicked spin-1/2 chains \cite{prosen02}, the kicked
quantum top \cite{manderfeld01}, Loschmidt echoes \cite{benenti03},
and experimentally for the imitation of the single particle quantum
problem in microwave billiards \cite{pance00}. In all these cases,
the quantum systems had one or several of the following features:
(i)  very few degrees of freedom; (ii) proximity to the classically
chaotic limit; (iii) application of external time-dependent forces,
removing the difficulty associated with the discrete frequency
spectrum of an isolated quantum system. In contrast, we deal here
with an essentially isolated system having a macroscopic number of
maximally non-classical components (spins 1/2), and have no
specially introduced precondition for chaos apart from the naturally
occurring non-integrable interaction between spins.

We note a remarkable fact revealed by Fig.~\ref{fig:decays}a: the
phases of the long-time oscillations of the 1.9~ms and 2.5~ms echoes
nearly coincide with each other and are shifted by $\pi$ with
respect to the FID phase. Indeed, one can observe that the
zero-crossings (cusps) of the FID and the two echoes coincide in the
long-time regime. Given that these are the absolute-value plots, the
above coincidences imply that the relative phases of the long-time
signals are either zero or $\pi$. These two possibilities can be
discriminated by keeping track of the successive sign changes at the
zero-crossings for each signal. (The inset of Fig.~\ref{fig:decays}a
shows the sign of the FID and each echo.) This may be a fundamental
phase relation associated with the fact that the 1.9 ms and 2.5 ms
echoes start after the FID has begun to approach the asymptotic
regime of Eq.~(\ref{eq:lt}). In contrast, the 0.56~ms echo starts
well before the FID has reached that regime, and its phase has no
particular relation to the other three signals.

We have observed a universal long-time behavior of $^{129}$Xe FIDs
and solid echoes in solid xenon. In all cases, a sinusoidally
modulated exponential decay sets in after just a few times $T_2$.
This behavior is universal in the sense that the two parameters
characterizing the long-time decay are independent of the NMR
pulse/delay sequence, even though each such sequence generates a
different initial spin configuration. These findings reveal a
fundamental property of nuclear spin dynamics. In addition, they
also support the idea that the correspondence between classical and
quantum chaotic properties of real many-body systems can be
established at the level of Pollicott-Ruelle resonances. Further
investigations, however, are required in order to clarify whether
the eigenmodes of form (\ref{eq:prsolution}) actually exist in
many-spin density matrices and, if so, how far this correspondence
can be taken.

The authors are grateful to M.~S. Conradi, T. Egami, K.~A. Lokshin,
and O.~A. Starykh for helpful discussions. This work was supported
by the National Science Foundation (PHY-0134980).

\

\

\appendix*

\noindent {\bf SUPPORTING MATERIAL}

\noindent{(with a separate system of references)}

\

\noindent {\bf Detailed Experimental Methods}

\

The samples of solid polycrystalline hyperpolarized $^{129}$Xe were
prepared using spin-exchange convection cells \cite{su04}. These
cells allow the production of large quantities of hyperpolarized
$^{129}$Xe by keeping most of the xenon in the liquid phase. The
xenon gas is polarized at 100~$^\circ$C and travels by convection to
the column of liquid that is kept cold ($-110$~$^\circ$C) in a
retort that is open to the rest of the cell. The liquid $^{129}$Xe
is polarized by passive phase exchange with the gas. Using these
$10$ cm$^{3}$ cells, we have generated a nuclear spin polarization
of 5-10\% in one liquid sample of $0.7$ mM enriched xenon ($86\%$
$^{129}$Xe, $ 0.13\%$ $^{131}$Xe ) and two liquid samples of $2.6$
mM natural xenon ($26.4\%$ $^{129}$Xe, $21.29\%$ $^{131}$Xe) inside
a $1.5$~T horizontal-bore superconducting magnet ($^{129}$Xe Larmor
frequency of $17.6$ MHz).

To study the NMR signals in the solid state, the liquid
hyperpolarized xenon was subsequently frozen and maintained at
$77$~K while still inside the magnet by forcing liquid nitrogen
directly into the chamber containing the cell. At 1.5~T, the
polarization is expected to survive this phase transition intact
\cite{kuzma02}, and this was verified experimentally by monitoring
the NMR signal height with small-angle pulses during freezing. Since
90$^{\circ}$ rf excitation pulses (pulse length $\approx 10$~$\mu$s)
are used to obtain the FIDs and solid echoes, the magnetization was
essentially destroyed by each acquisition. However, the convection
cells allow easy {\it in situ} regeneration of the $^{129}$Xe
polarization, and the experiment could thus be repeated many times
per day. The NMR flip angle was calibrated for each separate
measurement by applying several identical excitation pulses ($\theta
\approx 40^{\circ}$) and measuring the corresponding $\cos\theta$
magnetization loss. The separate acquisitions of the initial and
long-time portions of the decays for each FID and echo were joined
together at one point common to both signals and
amplitude-renormalized. (The renormalization is necessary because
the $^{129}$Xe polarization obtained by spin-exchange optical
pumping varies somewhat from run to run.) In one case (the 0.56~ms
delay time in enriched Xe), three separate signals had to be
acquired, and these were similarly joined together at two locations.

We have examined the possible effects of radiation damping and
inhomogeneity in the external field on the FID and solid echo line
shapes. The observed transverse decay time $T_2^{*}$ for
hyperpolarized liquid xenon (having approximately the same
polarization as the solid) is $\approx 23$~ms in the same NMR probe
and location in the magnet. (Here, $T_2^{*}$ refers to the timescale
for transverse decay from all sources of dephasing combined.) Unlike
the solid case, the spin-spin interactions in the motionally
narrowed liquid are much weaker, allowing the decay to be
essentially limited by some combination of radiation damping and
field inhomogeneity. (Since we observe the liquid lineshape to be
independent of polarization, it is likely that field inhomogeneity
is actually the limiting factor.) Since the liquid decay is much
longer than the $\approx 1$~ms decay time observed for the solid, we
conclude that the effects of these other sources of dephasing are
negligible in the solid. As an additional check, we note that the
decay shapes associated with both the field inhomogeneity and the
radiation damping are expected to be non-exponential; if these
factors were not negligible in our experiments, then the long-time
decay of the beat amplitude would exhibit a noticeable departure
from an exponential shape.

\

\

\noindent {\bf Theoretical facts about solid echoes}

\

Solid echoes are produced by the radio frequency pulse sequence
$90^{\circ}_y - \hbox{time delay} \, \tau - 90^{\circ}_x$. Here and
below all axes are defined in the Larmor rotating reference frame.
The pulses are assumed to rotate the spin system instantaneously.

Before the first pulse, the spin system is characterized by the
equilibrium density matrix
\begin{equation}
\rho_{\hbox{\scriptsize eq}} = C_N \, \hbox{exp} \left\{ {M_z H
\over k_B T} \right\}, \label{rhoeq}
\end{equation}
where $M_z = g \sum_k I_k^z$ is the total magnetization of nuclei
with gyromagnetic ratio $g$ in external magnetic field $H$ directed
along the $z$-axis, $k_B$ is Boltzmann's constant, $T$ the
temperature (in our case, it is the effective spin temperature of
the hyperpolarized system), and $C_N$ is the normalization constant.
The usual energy hierarchy $k_B T \, \gg \,  \hbar g H \, \gg \,
\hbar/T_2$ is satisfied in our experiment.

The operator of the first ($90^{\circ}_y$) pulse is
\begin{equation}
P_1 = \hbox{exp} \left\{ - i {\pi \over 2} \sum_k I_k^y \right\} .
\label{P1}
\end{equation}
This operator transforms $I_k^z \rightarrow P^{+}_1 I_k^z P_1 =
I_k^x$, and,  therefore, just after the pulse
\begin{equation}
 \rho(0) = P^{+}_1 \rho_{\hbox{\scriptsize eq}} P_1 =
C_N \, \hbox{exp} \left\{ {M_x H \over k_B T} \right\} \approx C_N
\left( \mathbbm{1} + {M_x H \over k_B T} \right), \label{rho0}
\end{equation}
where $M_x = g \sum_k I_k^x $.

Following the first pulse, the system undergoes free induction decay
(FID) described as follows:
\begin{eqnarray}
 F_{\hbox{\tiny FID}}(t) 
&
\cong \langle M_x(t) \rangle
 = \hbox{Tr} \left\{ \sum_k I_k^x
 \, e^{i {\mathcal{H}} \:  t} \, \rho(0) \, e^{-i {\mathcal{H}} \:  t}
\right\} 
\nonumber
\\
& 
\cong \hbox{Tr} \left\{ \sum_k I_k^x
 \, e^{i {\mathcal{H}} \:   t} \, \sum_n I_k^n \, e^{-i {\mathcal{H}}  \:  t}
\right\}, 
\label{FFID}
\end{eqnarray}
where the second sign $\cong$ implies that we expanded $\rho(0)$
according to Eq.(\ref{rho0}), dropped the term containing
$\mathbbm{1}$ as it makes no contribution to the trace, and then
omitted the prefactor ${C_N g^2 \over k_B T}$. The Hamiltonian
${\mathcal{H}}$ is given by Eq.(2) of the main article.

The second ($90^{\circ}_x$) pulse is characterized by operator
\begin{equation}
P_2 = \hbox{exp} \left\{ - i {\pi \over 2} \sum_k I_k^x \right\} .
\label{P2}
\end{equation}
The action of this operator is the following:
\begin{equation}
\begin{array}{rcl}
I_k^x \rightarrow & P^{+}_2 I_k^x P_2  & = I_k^x
\\
I_k^y \rightarrow & P^{+}_2 I_k^y P_2  & = - I_k^z
\\
 I_k^z \rightarrow & P^{+}_2 I_k^z P_2  & = I_k^y
\end{array}
\label{P2rules}
\end{equation}

The density matrix just before the second pulse is
\begin{equation}
 \rho_-(\tau) = e^{i {\mathcal H} \tau} \, \rho(0) \, e^{-i {\mathcal H} \tau }.
\label{rho-}
\end{equation}
Just after the pulse, it becomes
\begin{equation}
 \rho_+(\tau) =
P^{+}_2 \, e^{i {\mathcal H} \tau} \, \rho(0) \, e^{-i {\mathcal H}
\tau } \, P_2. \label{rho+}
\end{equation}
The action of $ P^{+}_2 \, ... \, P_2$ in Eq.(\ref{rho+}) can be
represented as the change of the identity of $I_k^{\alpha}$
operators according to rules (\ref{P2rules}). (Here $\alpha =
x,y,z$.)  Indeed, since $ P_2 P^{+}_2 = \mathbbm{1} $, any term of
the form $ P^{+}_2 \, I_k^{\alpha} I_l^{\beta}  ... I_n^{\delta} \,
P_2$ arising in the expansion of the right-hand-side of
Eq.(\ref{rho+}) is equal to $ P^{+}_2 \, I_k^{\alpha} \, P_2 P^{+}_2
\, I_l^{\beta} \, P_2 P^{+}_2 ... P_2 P^{+}_2 \, I_n^{\delta} \,
P_2$, and hence each operator $I_k^{\alpha}$ is being ``rotated''
individually. Using the above fact, Eq.(\ref{rho+}) can be rewritten
as
\begin{equation}
 \rho_+(\tau) =
e^{i {\mathcal H}_{\hbox{\tiny R}} \: \tau} \, \rho(0) \, e^{-i
{\mathcal H}_{\hbox{\tiny R}} \: \tau}, \label{rho+1}
\end{equation}
where
\begin{equation}
 {\mathcal H}_{\hbox{\tiny R}} = \sum_{k<n} B_{kn} I_k^x I_n^x  \, + \,
 A_{kn} I_k^y I_n^y  \, + \, B_{kn} I_k^z I_n^z .
\label{HR}
\end{equation}
(The subscript ``R'' stands for ``Rotated''.)

The density matrix at time $t^{\prime}$ after the second pulse is
\begin{equation}
 \rho_{\hbox{\tiny SE}}(\tau, t^{\prime}) = e^{i {\mathcal H} \:  t^{\prime}}
 \, \rho_+(\tau) \, e^{-i {\mathcal H} \:  t^{\prime} }.
\label{rhoSE}
\end{equation}
 Finally, the solid echo signal is given by
\begin{equation}
 F_{\hbox{\tiny SE}}(\tau, t^{\prime}) \cong
\hbox{Tr} \left\{ \sum_k I_k^x
 \; \; e^{i {\mathcal H} \:  t^{\prime}} \,
e^{i {\mathcal H}_{\hbox{\tiny R}} \:  \tau} \, \sum_n I_k^n \: e^{-
i {\mathcal H}_{\hbox{\tiny R}} \:  \tau} \, e^{-i {\mathcal H} \:
t^{\prime}} \right\}. \label{FSE}
\end{equation}

As mentioned in the main article, different delay times $\tau$ imply
fundamentally different initial spin configurations, $\rho_+(\tau)$,
for the solid echo response. At small $\tau$, $\rho_+(\tau)$ is not
much different from $\rho(0)$ given by Eq.(\ref{rho0}), which
describes uncorrelated uniformly polarized spin distribution. As
$\tau$ increases, the density matrix $\rho_+(\tau)$ is gradually
overtaken by increasingly higher order spin correlations induced by
the Hamiltonian ${\mathcal H}_{\hbox{\tiny R}}$. This can be seen
from the expansion
\begin{eqnarray}
 \rho_+(\tau) = 
& 
\rho(0)
+ i \tau [{\mathcal H}_{\hbox{\tiny R}}, \sum_k I_k^x] - {\tau^2
\over 2!} [{\mathcal H}_{\hbox{\tiny R}},[{\mathcal H}_{\hbox{\tiny
R}}, \sum_k I_k^x]] 
\nonumber
\\
&
- {i \tau^3 \over 3!} [{\mathcal H}_{\hbox{\tiny
R}},[{\mathcal H}_{\hbox{\tiny R}},[{\mathcal H}_{\hbox{\tiny R}},
\sum_k I_k^x]]] + ... \; \; . \label{rho+exp}
\end{eqnarray}
Each commutation with ${\mathcal H}_{\hbox{\tiny R}}$ entails an
extra power of operators $I_k^{\alpha}$. The resulting higher order
spin correlations dominate in $\rho_+(\tau)$ at $\tau \gtrsim T_2$.
We further note that in this regime, the initial spin configurations
for a solid echoes with different delay times $\tau$ are not
equivalent to each other.

The correlations induced by ${\mathcal H}_{\hbox{\tiny R}}$ are
non-trivially connected with those intrinsic for ${\mathcal H}$, the
actual Hamiltonian of the system. One can achieve a limited
simplification of the problem by expressing ${\mathcal
H}_{\hbox{\tiny R}}$ as
\begin{equation}
 {\mathcal H}_{\hbox{\tiny R}} = - {\mathcal H} + \Delta {\mathcal H},
\label{HRDeltaH}
\end{equation}
where
\begin{equation}
 \Delta {\mathcal H} = \sum_{k<n} 2 B_{kn} I_k^x I_n^x  \, + \,
 (A_{kn} + B_{kn}) (I_k^y I_n^y  \, + \, I_k^z I_n^z) .
\label{DeltaH}
\end{equation}
The advantage of this representation is that $\Delta {\mathcal H}$
commutes with $\sum_k I_k^x$. If it also commuted with ${\mathcal
H}$, then the exponent $e^{i (- {\mathcal H} + \Delta {\mathcal H})
\: \tau} $ would factorize into $e^{- i {\mathcal H} \: \tau} \;
e^{i \Delta {\mathcal H} \: \tau}$, and then $e^{i \Delta {\mathcal
H} \: \tau}$ would not influence $\rho_+(\tau)$, while $e^{- i
{\mathcal H} \: \tau}$ would lead to a time-reversed evolution and
imply the full recovery of the initial state at $t^{\prime} = \tau$
[i.e.  $\rho_{\hbox{\tiny SE}}(\tau, \tau) = \rho(0)$].

As mentioned in the main article, such a situation is, indeed,
realized for a pair of interacting spins 1/2. In this case,
$[{\mathcal H}, \Delta {\mathcal H}] = 0$, because $[I_1^{\alpha}
I_2^{\alpha}, \; I_1^{\beta} I_2^{\beta}]=0$ for any $\alpha$ and
$\beta$. However, for a lattice of spins 1/2, $[{\mathcal H}, \Delta
{\mathcal H}] \neq 0$, because the commutators like $[I_1^{\alpha}
I_2^{\alpha}, \; I_2^{\beta} I_3^{\beta}]$ are not equal to zero for
$\alpha \neq \beta$.

One can nevertheless conclude that solid echo tends to undo two-spin
correlations in many-spin systems. Therefore,  at sufficiently small
$\tau$, when $\rho_+(\tau)$ is controlled by the first three terms
in expansion (\ref{rho+exp}), the solid echo can be described as an
imperfect attempt at time reversal. (Note that even for small delay
time 0.56~ms, the maximum of the solid echo in Fig. 2a of the main
article occurs noticeably earlier than at $t = 2\tau$.) When $\tau
\gtrsim T_2$, the rigorous calculation of solid echoes becomes
intractable, but otherwise it is clear the concept of time reversal
becomes increasingly inappropriate.

For further reading, see the original references concerning solid
echoes \cite{powles62,powles63}.

\

\

\newpage
\noindent {\bf Natural Xenon Data} \vspace{0.5in}
\renewcommand{\thefigure}{S1}
\begin{figure}[h]
\begin{center}
\includegraphics[width=3.2in]{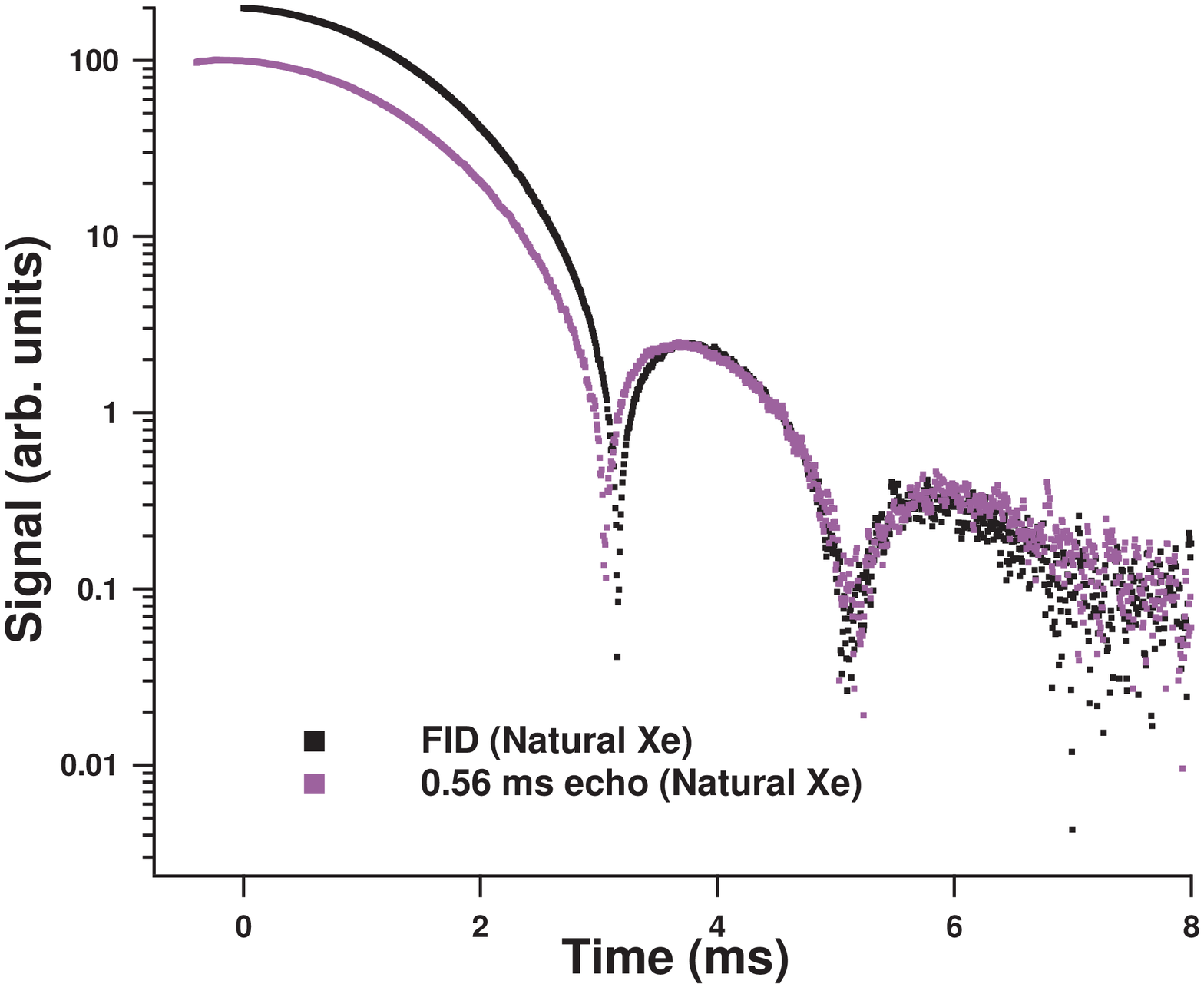}
\caption{\label{fig:natxe} Representative acquisitions of FID and
the 0.56~ms solid echo recorded for $^{129}$Xe in polycrystalline
xenon with natural abundance of nuclear isotopes at 77 K. Here, the
FID and echo and normalized and time-shifted in the same manner as
for enriched xenon in Fig.~2b (main manuscript) to show the overlap
in the long-time regime. The long-time behavior of the FID and solid
echo are identical in form, although the decay coefficient $\gamma$
and beat frequency $\omega$ each have values smaller than those for
enriched xenon; see Table~I of main manuscript.}
\end{center}
\end{figure}


\begin{thebibliography}{99}

\bibitem{gaspardbook}
P. Gaspard, {\it Chaos, Scattering and Statistical Mechanics\/}
(Cambridge University Press, Cambridge, 1998).

\bibitem{haakebook}
F. Haake, {\it Quantum Signatures of Chaos\/} (Springer-Verlag,
Berlin, Germany, 2001).

\bibitem{gaspard98}
P.~Gaspard, M.~E.~Briggs, M.~K.~Francis, J.~Sengers, R.~W.~Gammon,
J.~R~Dorfman, and R.~V.~Calabrese, Nature {\bf 394}, 865 (1998).

\bibitem{dettmann99}
C.~P.~Dettmann, E.~G.~D.~Cohen, and H.~van~Beijeren, Nature {\bf
401}, 875 (1999).

\bibitem{grassberger99}
P.~Grassberger and T.~Schreiber, Nature {\bf 401}, 875 (1999).

\bibitem{gaspard99}
P.~Gaspard, M.~E.~Briggs, M.~K.~Francis, J.~Sengers, R.~W.~Gammon,
J.~R~Dorfman and R.~V.~Calabrese, Nature {\bf 401}, 876 (1999).

\bibitem{fine05}
B.~V.~Fine, Phys. Rev. Lett. {\bf 94}, 247601 (2005).

\bibitem{walker97}
T.~G.~Walker and W.~Happer, Rev. Mod. Phys. {\bf 69}, 629 (1997).

\bibitem{gatzke93} M. Gatzke, G.~D.~Cates, B. Driehuys, D. Fox, W.
Happer, and B. Saam, Phys. Rev. Lett. {\bf 70}, 690 (1993).

\bibitem{yen63} W.~M.~Yen and R.~E.~Norberg, Phys. Rev. {\bf 131},
269 (1963).

\bibitem{lowe57}
I.~J.~Lowe and R.~E.~Norberg, Phys. Rev. {\bf 107}, 46 (1957).

\bibitem{fine04}
B.~V.~Fine, Int. J. Mod. Phys. B {\bf 18}, 1119 (2004).

\bibitem{fine97}
B.~V.~Fine, Phys. Rev. Lett. {\bf 79}, 4673 (1997).

\bibitem{epaps} See Supporting Material at the end of this paper for detailed experimental methods, a theoretical description of solid echoes, and natural
xenon data. 

\bibitem{powles62}
J.~G.~Powles and P.~Mansfield, Phys. Lett. {\bf 2}, 58 (1962).

\bibitem{hahn50}
E.~L.~Hahn, Phys. Rev. {\bf 80}, 580 (1950).

\bibitem{rhim71}
W.~K.~Rhim, A.~Pines and J.~S.~Waugh, Phys. Rev. B {\bf 3}, 684
(1971).

\bibitem{cho05}
H.~Cho, T.~D.~Ladd, J.~Baugh, D.~G.~Cory and C.~Ramanathan, Phys.
Rev. B {\bf 72}, 054427 (2005).

\bibitem{slichterbook}
C.~P.~Slichter,  {\it Principles of Magnetic Resonance\/}
(Springer-Verlag, New York, NY, 1996 corr. ed.).

\bibitem{abragam61}
A.~Abragam, {\it Principles of Nuclear Magnetism\/} (Oxford Science
Publications, New York, NY, 1961).

\bibitem{vvleck48}
J.~H.~Van~Vleck, Phys. Rev. {\bf 74}, 1168 (1948).

\bibitem{klauder62}
J.~R. Klauder and P.~W.~Anderson, Phys.Rev. {\bf 125}, 912 (1962).

\bibitem{powles63}
J.~G.~Powles and J.~H.~Strange, Proc. Phys. Soc. {\bf 82}, 6 (1963).

\bibitem{borckmans68}
P.~Borckmans and  D.~Walgraef,   Phys. Rev. {\bf 167}, 282-288
(1968).

\bibitem{cowanbook}
B.~Cowan, {\it Nuclear Magnetic Resonance and Relaxation\/}
(Cambridge University Press, Cambridge, 1997).

\bibitem{enge74}
M.~Engelsberg and I.~J.~Lowe, Phys. Rev. B {\bf 10}, 822 (1974).

\bibitem{fine03}
B.~V.~Fine, J. Stat. Phys. {\bf 112}, 319 (2003).

\bibitem{fabricius97}
K.~Fabricius, U.~L{\"{o}}w and J.~Stolze, Phys. Rev. B {\bf 55},
5833 (1997).

\bibitem{su04}
T.~Su, G.~L.~Samuelson, S.~W.~Morgan, G.~Laicher and B.~Saam, Appl.
Phys. Lett. {\bf 85}, 2429 (2004).

\bibitem{ruelle86}
D.~Ruelle, Phys. Rev. Lett. {\bf 56}, 405 (1986).

\bibitem{prosen02}
T. Prosen, J. Phys. A {\bf 35}, L737 (2002).

\bibitem{manderfeld01}
C.~Manderfeld, J.~Weber and F.~Haake, J. Phys. A {\bf 34}, 9893
(2001).

\bibitem{benenti03}
G.~Benenti, G.~Casati, and G.~Veble, Phys. Rev. E {\bf 67},
055202(R) (2003).

\bibitem{pance00}
K.~Pance, W.~Lu, and S.~Sridhar, Phys. Rev. Lett. {\bf 85}, 2737
(2000).

\end{thebibliography}

\begin{thebibliography}{9}

\bibitem{su04} T. Su, G. L. Samuelson, S. W. Morgan, G. Laicher, and B. Saam,
Appl. Phys. Lett. {\bf 85}, 2429 (2004).

\bibitem{kuzma02}
N. N. Kuzma, B. Patton, K. Raman, and W. Happer, Phys. Rev. Lett.
{\bf 88}, 147602 (2002).

\bibitem{powles62} J. G. Powles and P. Mansfield, Phys. Lett. {\bf 2}, 58
(1962).

\bibitem{powles63} J. G. Powles and J. H. Strange, Proc. Phys. Soc. {\bf 82}, 6
(1963).

\end{thebibliography}
\end{document}